# Ultra-Clean Freestanding Graphene by Platinum-Metal Catalysis

*Jean-Nicolas Longchamp\*, Conrad Escher, Hans-Werner Fink*

*Physics Institute, University of Zurich, Winterthurerstrasse 190, 8057 Zurich, Switzerland*

**\*Corresponding Author**

E-mail: longchamp@physik.uzh.ch

**ABSTRACT**

While freestanding clean graphene is essential for various applications, existing technologies for removing the polymer layer after transfer of graphene to the desired substrate still leave significant contaminations behind. We discovered a method for preparing ultra-clean freestanding graphene utilizing the catalytic properties of platinum metals. Complete catalytic removal of polymer residues requires annealing in air at a temperature between 175 and 350°C. Low-energy electron holography investigations prove that this method results in ultra-clean freestanding graphene.

**TEXT**

The physical and electronic properties of graphene[1, 2] depend to a large extent on its defect-free structure and its cleanliness. Scattering of transport electrons at impurities is one of the major drawbacks in the use of graphene in electronic devices[3-5]. When employing graphene as a substrate in electron microscopy, the presence of residues is frustrating because these features are often of the same size as the object under study[6]. While the growth of defect-free single-layer graphene by means of chemical vapor deposition (CVD) is nowadays routinely possible[7, 8], easily accessible and reliable techniques to transfer graphene to different substrates in a clean manner are still lacking. The common technique for the transfer of the layers grown by means of CVD on a metallic substrate (usually nickel or copper) onto an arbitrary substrate is based on the use of a polymer layer, ordinarily polymethyl methacrylate (PMMA), spread or spin-coated over graphene[5, 9, 10]. The removal of the approximately 100nanometer thick PMMA layer is a challenge and extensive efforts have been undertaken in the past few years to establish a reliable technique to retrieve the pristine graphene without PMMA residues[3-5, 11-14]. Well-known chemical etchants for PMMA are acetone and chloroform[15]. Unfortunately, wet chemical treatment of the polymer leads to contaminated graphene layers with lots of residues left behind. Thermal annealing at temperature in the range of 300-400°C in vacuum [12, 15] or in an Ar/$H_2$ atmosphere[15] appears to help the cleaning process. However, besides the fact that these techniques are not easily available, they do not lead to ultra-clean freestanding graphene.

We present here a simple method for preparing ultra-clean freestanding graphene based on the removal of a polymer layer by the catalytic activity of platinum metals (such as Pt, Pd, Rh). The cleanliness of the prepared freestanding graphene layers is investigated by means of low-energy electron holography[16, 17]. Electrons with kinetic energy in the range of 50-100eV are extremely sensitive to the smallest amounts of contamination causing electrical field disturbances originating, for instance, from electrical non-conductive PMMA residues. Furthermore, low-energy electrons exhibit a scattering cross-section for atoms almost independent of their Z number, an advantage in comparison to TEM in view of detecting the presence of possible hydrocarbon residues.

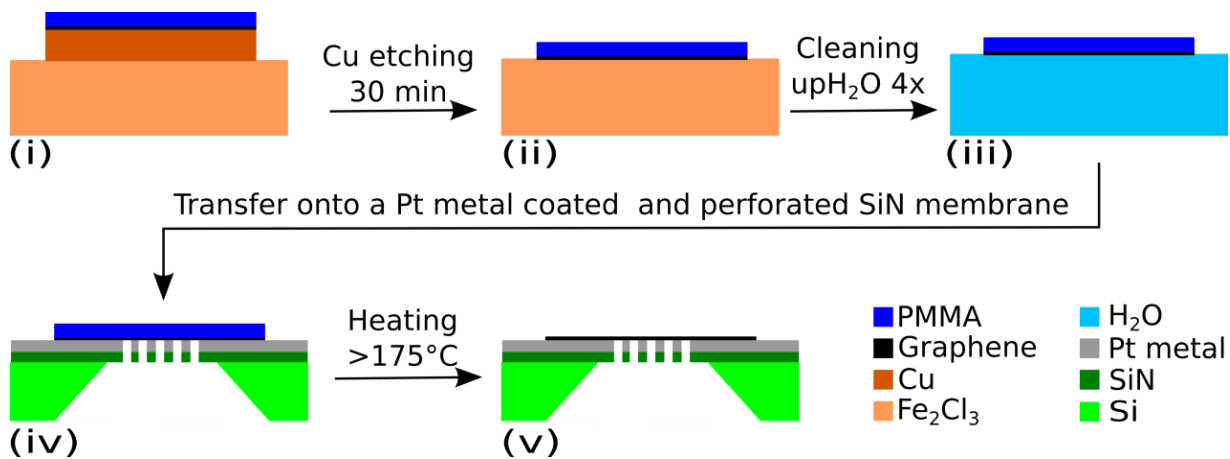

**Figure 1:** Flow chart of the preparation of ultra-clean freestanding graphene.

All the graphene layers used in this study were grown by conventional CVD method[18] on a polycrystalline copper substrate. A PMMA capping (about 100nm thick) is subsequently spin-coated onto the graphene (Fig. 1(i)). The preparation of the clean freestanding graphene starts with the chemical wet-etching of the metallic substrate, as illustrated in Fig. 1(ii). After the complete removal of the underlying copper, the remaining graphene-PMMA complex is rinsed four times with ultra-purified water to wash off the etching solution (Fig. 1(iii)). The third step of the preparation method consists of the placement of the graphene-PMMA composite on a metal (Pt, Pd or Rh)-coated 50nm thick silicon nitride membrane previously perforated by means of a focused gallium ion beam (Fig.1 (iv)). After drying, the sample is placed onto a conventional laboratory heating plate for thermal annealing in air at a temperature in the range of 175 to 350°C (Fig. 1(v)); i.e. well below the oxidation temperature of graphene in air[12, 19].

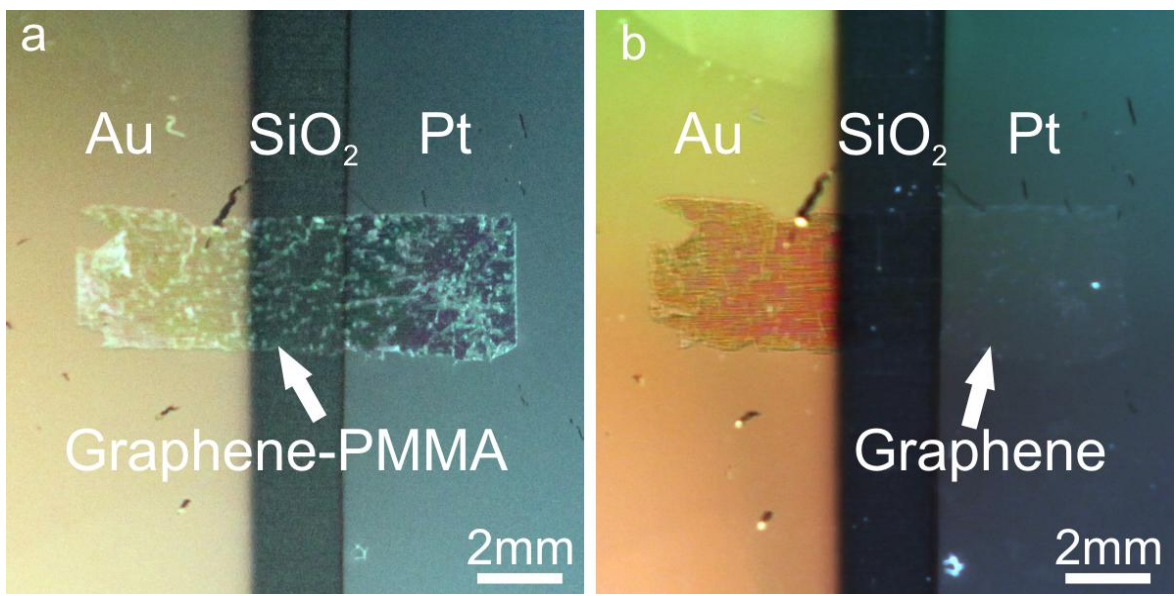

**Figure 2:** a. Optical photograph of a graphene-PMMA composite transferred onto a glass plate covered with gold and platinum before thermal annealing. b. Optical photograph of the same sample after thermal annealing at 300°C for 30min.

Figure 2 illustrates the degradation of the PMMA capping due to the presence of the catalytic metal while it remains present on a noble metal, i.e., gold in this case. On a 10x10mm microscopy glass plate, about half of its surface is coated with gold and the other half with platinum. A 2mm gap of uncoated $SiO_2$ between the two metal layers was arranged in order to avoid cross-diffusion of the metals. Figure 2(a) displays the situation after the transfer of a graphene-PMMA composite onto the glass plate and across the gap between the gold and the platinum layers prior to thermal annealing. The PMMA is clearly visible on both metals. Figure 2(b) shows the result after the annealing of the glass plate at 300°C for 30min. The PMMA was decomposed above the platinum layer, leaving clean graphene on the catalytic active metal. Similar results have been obtained for annealing at temperatures in the range of 175-350°C, provided that the annealing time is properly adjusted. It varies between 6h for a temperature of 175°C and just 3min at 350°C. As evident from Figure 2, the PMMA remains present on the gold surface and is apparently un-degraded.

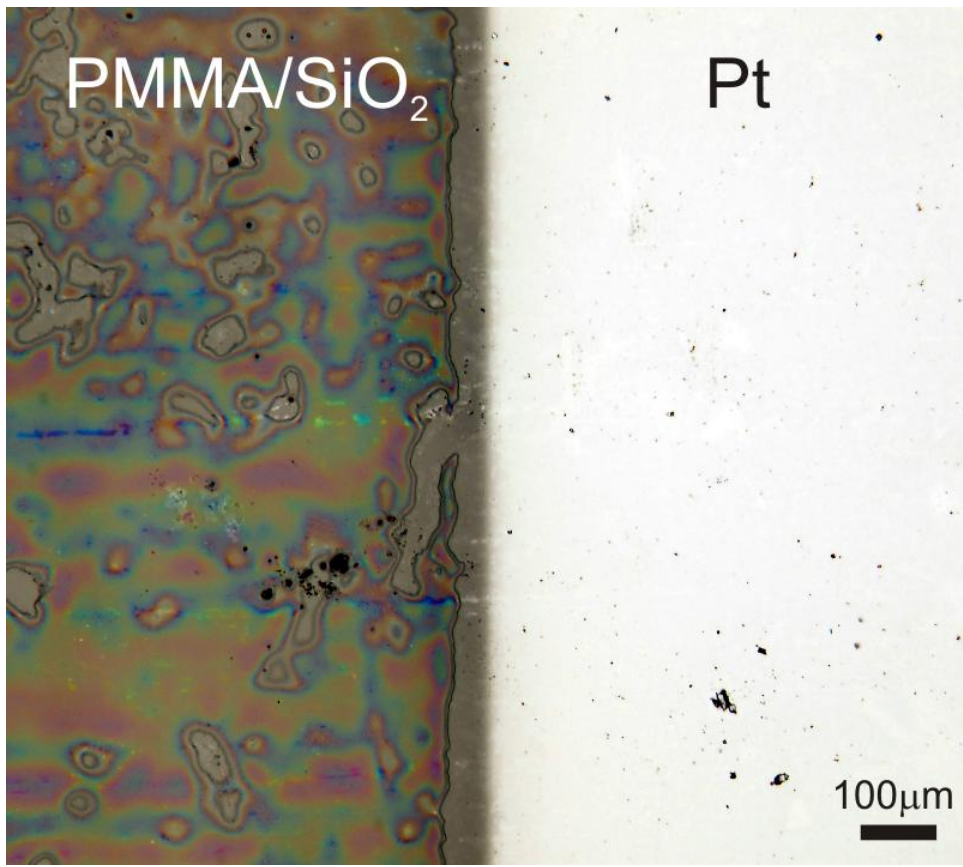

**Figure 3:** Light optical microscopy image of the sample shown in Figure 2b.

Figure 3 shows a light optical microscopy image of the sample presented in Figure 2, but at higher magnification. As mentioned above, after annealing the PMMA is degraded on the Pt surface while it remains intact on almost the entire $SiO_2$ surface; a characteristic interference contrast originating from the presence of PMMA on the glass is clearly visible. An important aspect to note is that a region extending as far as 80micron from the Pt-metal coating out to the bare $SiO_2$ was also cleaned by the thermal annealing process. It appears that the proximity of platinum is already sufficient to cause a catalytic reaction leading to the decomposition of PMMA. These findings clearly justify the hope that an extended region of clean freestanding graphene can be obtained by the method described above.

The thermal degradation of polymers is a complicated process and subject of an entire research field[20-22]. Various decomposition mechanisms have been proposed, such as random-chain scission, end-chain initiation, unzipping, depropagation or depolymerization, to name just a few. For PMMA in particular, the end-chain initiation has been established as the predominant process[20]. A common aspect in all the different decomposition paths is the involvement of hydrogen in the process. Normally, during the thermal degradation of polymers, hydrogen originates from either the backbone or the side chains of the macromolecule, resulting in the formation of smaller molecules or the degradation into monomers. However, this process requires high temperatures, at least 400°C for PMMA[20]. If the molecules produced in this process are small enough, they can escape into the gas phase. In the case presented here, the catalytic aspect of the process can be explained as hydrogenation of the PMMA promoted by the platinum metals such that the cracking of the polymer occurs already at much lower temperatures. The platinum metals are well known in catalysis. It is conceivable that the reaction is initiated by the ability of the Pt-metal to dissociate molecular adsorbed $H_2$ into atomic hydrogen. The fact that a minimal temperature must be attained to start such a reaction (175°C in our case), that the time for completing the reaction decreases rapidly with increasing temperature, and that with other metals such as gold the reaction does not proceed at all, are additional strong indications for the catalytic character of the process described here.

In order to investigate the cleanliness of such graphene samples on the nanometer-scale, we prepared freestanding graphene layers placed over holes between 250 and 1000nm in diameter milled by a focused gallium ion beam into a Pt-coated SiN membrane. In Figure 4(a) a SEM image of such a hole and in (b) a high magnification image of the substrate surface after sputter deposition of Pt are displayed. The thickness of the platinum layer amounts to 15nm. It is evident that the metal layer is not uniform but exhibits islands of about 50nm in size. It is believed that these nanometer-sized domains promote the catalytic activity of the platinum layer. This is confirmed by the use of smoother Pt layers obtained by e-beam evaporation which leads to an increase of the catalytic reaction time to 45min at 300°C in contrast to just 5min in case of rough sputtered layers.

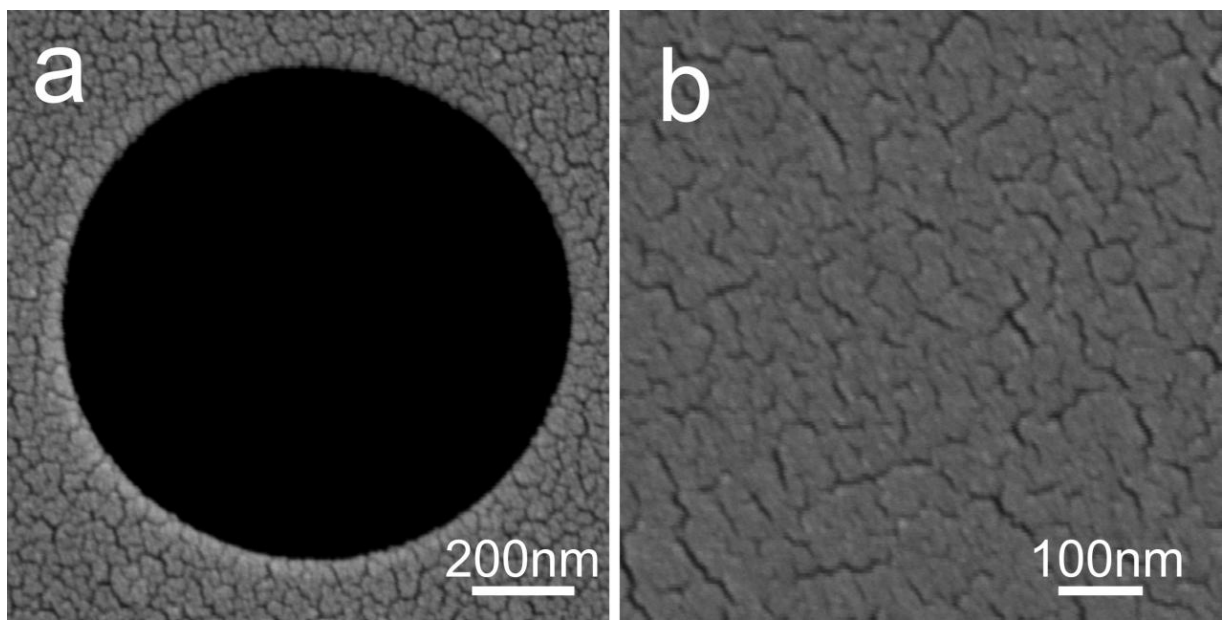

**Figure 4:** a. Scanning electron image of a hole of 1000nm in diameter milled in a SiN membrane and subsequently coated with Pt. b. Higher magnification image of the clustered Pt layer.

After the thermal annealing process, the prepared samples were directly transferred to our low-energy electron point source microscope. In this holographic setup inspired by the original idea of Gabor[23] of in-line holography, a sharp (111) -oriented tungsten tip acts as source of a divergent beam of highly coherent electrons[24]. The electron emitter can be brought as close as 200nm to the sample with the help of a 3-axis piezo-manipulator. Part of the electron wave impinging onto the sample is elastically scattered and represents the object wave, while the un-scattered part of the wave represents the reference wave[16]. At a distant detector, the interference pattern between the object wave and the reference wave – the hologram – is recorded. The magnification in the image is given by the ratio of detector-tip-distance to sample-tip-distance and can be as high as one million. Figure 5(a) displays an image of a freestanding ultra-clean graphene layer covering a 500nm diameter hole recorded with our low-energy electron point source microscope at an electron kinetic energy of 61eV and a total electron current of 50nA. Apparently, freestanding clean graphene is almost transparent even for low-energy electrons[6, 25]. The presence of graphene can only be confirmed by observing individual absorbates, possibly from the gas phase, sticking to the monolayer. For comparison, Figure 5(b) shows an image of freestanding graphene (70eV, 500nA), where the polymer layer was removed in the common manner by dissolving it in acetone. The resulting graphene layer is still polluted and almost opaque, even with a tenfold increased electron current, and the presence of PMMA residues is evident.

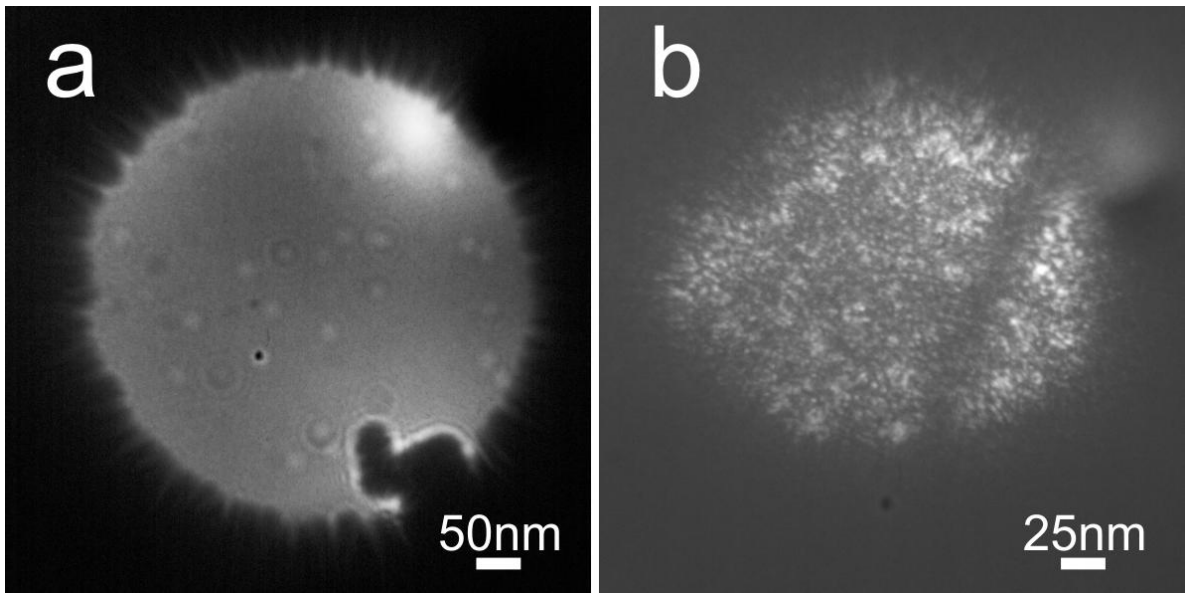

**Figure 5:** a. Low-energy electron hologram of ultra-clean freestanding graphene prepared by the method presented here. b. Electron transmission after the removal of the PMMA layer with acetone.

Similar results as to that presented in Figure 5(a) were also obtained with Pd as catalyst. Attempts to use other metals such as gold for the transfer and cleaning of graphene failed. Low-energy electron microscopy investigations revealed, in these cases, either empty holes where the graphene broke, or opaque holes where the graphene was heavily contaminated.

In summary, we have demonstrated that the use of platinum metals for the transfer of graphene leads to ultra-clean freestanding layers. The decomposition process of the PMMA layer is of catalytic nature and is promoted by the presence of a platinum metal. Even in the vicinity of the metal the polymer layer is removed, revealing clean graphene on an arbitrary substrate or even freestanding. The degradation reaction proceeds in air and at temperatures ranging from 175 to 350°C. This preparation method is thus easily accessible in every laboratory and does not require any special equipment. With this, ultra-clean graphene is now routinely available to serve not just as substrate for electron microscopy, but also in several applications needing ultra-clean graphene as pre-requisite; such as novel mechanical or electronic mesoscopic devices or as molecular sieves.

**ACKNOWLEDGMENT**

The authors are grateful for financial support by the Swiss National Science Foundation.